\documentstyle[aps,12pt]{revtex}
\newcommand{\bb}{\begin{equation}}
\newcommand{\ee}{\end{equation}}
\newcommand{\ba}{\begin{array}}
\newcommand{\ea}{\end{array}}

\begin {document}
\baselineskip 2.2pc

\title{Scattering of relativistic particles with
Aharonov--Bohm--Coulomb interaction in two dimensions\thanks
{published in J. Phys. A {\bf 33} (2000) 5049-5057.}}
\author{Qiong-gui Lin\thanks{E-mail:
        qg\_lin@163.net, stdp@zsu.edu.cn}}
\address{China Center of Advanced Science and Technology (World
	Laboratory),\\
        P.O.Box 8730, Beijing 100080, People's Republic of China\\
        and\\
        Department of Physics, Zhongshan University, Guangzhou
        510275,\\
        People's  Republic of China \thanks{Mailing address}}

\maketitle
\vfill

\begin{abstract}
{\normalsize The Aharonov--Bohm--Coulomb potentials in two
dimensions may describe the interaction between two particles
carrying electric charge and magnetic flux, say, Chern--Simons
solitons, or so called anyons. The scattering problem for such
two-body systems is extended to the relativistic case, and the
scattering amplitude is obtained as a partial wave series.
The electric charge and magnetic flux is ($-q$, $-\phi/Z$) for one
particle and ($Zq$, $\phi$) for the other.
When $(Zq^2/\hbar c)^2\ll 1$, and $q\phi/2\pi\hbar c$
takes on integer or half integer values,
the partial wave series is summed up approximately to give
a closed form. The results exhibit
some nonperturbative features and cannot be obtained from perturbative
quantum electrodynamics at the tree level.}
\end{abstract}
\vfill
\leftline {PACS number(s): 03.65.Pm; 03.65.Bz; 11.80.Et}
\leftline {Keywords: Relativistic scattering;
Aharonov--Bohm--Coulomb interaction; Two dimensions}
\newpage                                                    

\baselineskip 15pt

In a recent letter, we have studied the scattering of relativistic
electrons (or positrons) by the Coulomb field of a nucleus
in two dimensions [1]. The Dirac equation was solved in polar
coordinates, and the
scattering amplitude was obtained as a partial wave series.
For light nuclei the series can be summed up approximately
to give a closed result. 
The result, though being approximate, exhibits
some nonperturbative features and cannot be obtained from the
lowest-order contribution of perturbative quantum electrodynamics
(QED). This feature is not
manifest in  the corresponding result in three dimensions.
The purpose of the present paper is to extend our previous work
to the case with both Aharonov--Bohm (AB) and Coulomb potentials.
In the nonrelativistic case, this can be described by the
stationary Schr\"odinger equation
\bb
-{\hbar^2\over 2\mu}\left(\nabla+i{q\phi\over 2\pi\hbar c}
\nabla\theta\right)^2\psi-{Zq^2\over r}\psi=E\psi
\ee     
where $(r,\theta)$ are polar coordinates on the $xy$ plane,
and has been studied in the literature [2, 3]. Note that this
is a two-dimensional model and is different from the three-demensional
Aharonov--Bohm--Coulomb (ABC) system studied in the literature
[4-9]. Of course the AB potential is the same
in either two or three dimensions. The difference lies in the
Coulomb field. Previously the ABC system (in two or three
dimensions) is regarded as describing a charged particle (with
charge $-q$) moving in both an AB (with flux $\phi$) and a Coulomb
field generated by external sources.
Thus if the Coulomb field is generated by a nucleus
(with charge $Zq$), one should have a magnetic flux string
fixed on the nucleus to generate the AB potential. 
The situation seems difficult to realized in practice.
In a recent paper [3], we have shown that the above equation
can describe the relative motion of two particles, one carrying
electric charge and magnetic flux ($-q$, $-\phi/Z$) and the other
($Zq$, $\phi$), where $Z\ne 0$ is a real number.
Then $\mu$ is the reduced mass of the system.
Such particles appear in (2+1)-dimensional
Chern-Simons field theories as charged vortex soliton solutions
[10-17]. They are also called anyons because their angular momentum
may take on values other than integers or half integers.
Applications of such objects can be found in the study of fractional
quantum Hall effect [18-21], superconductivity [21-22], and repulsive
Bose gases [23].
Of course the solitons have finite sizes, and the real interaction
between them may be rather complicated. Thus the above description
by the two-dimensional ABC system is merely a rough approximation.
Anyway, the model may be of interest in itself since exact analysis
is possible. A relativistic case described by the Klein-Gordon
equation has been studied in the literature [24]. Here we will deal
with a spin $\frac 12$ particle described by the Dirac equation.
                     
Let us begin with the stationary Dirac equation translated from
Eq. (1):
\bb
\left[c{\bbox\alpha}\cdot\left({\bf p}+{q\phi\over 2\pi c}
\nabla\theta\right)+\gamma^0\mu c^2-{\kappa\over r}\right]\psi
=E\psi,
\ee     
where $\kappa=Zq^2$, ${\bf p}$ is the momentum operator,
${\bbox\alpha}=\gamma^0{\bbox\gamma}$, and
$\gamma^\mu=(\gamma^0, {\bbox\gamma})$ ($\mu=0,1,2$)
are Dirac matrices satisfying $\{\gamma^\mu, \gamma^\nu\}=2g^{\mu\nu}$
where $g^{\mu\nu}={\rm diag}(1,-1,-1)$. In two dimensions the Dirac
matrices can be realized by $2\times 2$ matrices:
\bb
\gamma^0=\sigma^3,\quad \gamma^1=i\sigma^1,\quad
\gamma^2=i\sigma^2,
\ee     
where the $\sigma$'s are Pauli matrices.
Thus $\psi$ is a two-component spinorial wave function.
As in the case with a pure Coulomb field, it is not difficult to
show that the conserved total angular momentum operator is
\bb
J=xp_y-yp_x+{i\hbar\over 2}\gamma^1\gamma^2.
\ee     
Therefore the particle has spin $\frac 12$.

The Dirac equation (2) can be solved in the polar coordinates by
separation of variables. Bound-state solutions (for $Z>0$ or
$\kappa>0$) are relatively easy to obtained, and some results will
be given at the end of the paper. Scattering solutions exist when
$E>\mu c^2$ or $E<-\mu c^2$. The latter correspond to antiparticles
after second quantization. At the level of single-particle theory,
their scattering can be treated formally in a way similar to the
former. Thus we only consider scattering solutions with $E>\mu c^2$
in the following. We use the representation (3). Let
\bb
{q\phi\over 2\pi\hbar c}=m_0+\nu,
\ee     
where $m_0$ is an integer and $-\frac 12<\nu\le \frac 12$, and
\bb
\psi_j(r,\theta)=\left(
\begin{array}{c}
f(r) \exp[i(j-m_0-1/2)\theta]/\sqrt r\\
g(r) \exp[i(j-m_0+1/2)\theta]/\sqrt r
\end{array}
\right),\quad
j=\pm\frac12, \pm\frac32, \ldots.
\ee     
It is easy to show that $J\psi_j=(j-m_0)\hbar\psi_j$.
Thus $j$ is a good quantum number. Substituting this expression
into Eq. (2) we obtain
two coupled ordinary differential equations for the radial wave
functions:
$$
{df\over dr}-{j+\nu\over r}f+k_1g+{\gamma\over r}g=0,
\eqno(7{\rm a})$$
$$
{dg\over dr}+{j+\nu\over r}g-k_2f-{\gamma\over r}f=0,
\eqno(7{\rm b})$$
where
\addtocounter{equation}{1}
\bb
k_1={E+\mu c^2\over\hbar c},\quad
k_2={E-\mu c^2\over\hbar c},\quad
\gamma={\kappa\over\hbar c}={Zq^2\over\hbar c}.
\ee     
Then we introduce the new variable
\bb
\rho=kr,\quad k=\sqrt{k_1k_2}={\sqrt{E^2-\mu^2 c^4}\over\hbar c},
\ee     
and two new functions $u(\rho)$, $v(\rho)$ through the definition
\bb
f(r)=\frac 12\sqrt{k_1} e^{i\rho}[u(\rho)+v(\rho)],\quad
g(r)=-\frac i2\sqrt{k_2} e^{i\rho}[u(\rho)-v(\rho)],
\ee     
to recast Eq. (7) into the form
$$
{du\over d\rho}-{i\beta\over\rho}u-{i\beta'+j+\nu\over\rho}v=0,
\eqno(11{\rm a})$$
$$
{dv\over d\rho}+2iv+{i\beta\over\rho}v+{i\beta'-j-\nu\over\rho}u=0,
\eqno(11{\rm b})$$
where
$$
\beta={\gamma\over 2}\left(\sqrt{k_1\over k_2}+\sqrt{k_2\over k_1}
\right)={\kappa\over\hbar v_{\rm c}},
\eqno(12{\rm a})$$
$$
\beta'={\gamma\over 2}\left(\sqrt{k_1\over k_2}-\sqrt{k_2\over k_1}
\right)=\beta\sqrt{1-{v_{\rm c}^2\over c^2}},
\eqno(12{\rm b})$$
where $v_{\rm c}$ is the classical velocity of the incident particle.
This is more convenient. Indeed, one can eliminate $v$ immediately to
obtain an equation for $u$ alone:
\addtocounter{equation}{2}
\bb
\rho{d^2u\over d\rho^2}+(1+2i\rho){du\over d\rho}+\left(2\beta-
{(j+\nu)^2-\gamma^2\over \rho}\right)u=0,
\ee     
where we have used $\beta^2-\beta'^2=\gamma^2$.
In this paper we assume for convenience that $|\gamma|<\frac 12$ (for
electron-nucleus interaction, this means that $Z\le 68$, which is in
general satisfied in practice). This is
not quite enough. If $|\nu|$ is not close to $\frac 12$, we further
assume that $|\gamma|<|\frac 12\pm \nu|$. (This cannot hold if $|\nu|$
is very close to $\frac 12$, which causes some difficulty and will be
discussed separately in the following. A larger $\gamma$ causes more
difficulty.) Then for any $j$, the solution of Eq. (13) is well
behaved at  the origin. Let
\bb
u(\rho)=\rho^s w(\rho),\quad s=\sqrt{(j+\nu)^2-\gamma^2}.
\ee     
Then we have for $w$ the equation
\bb
\rho{d^2w\over d\rho^2}+(2s+1+2i\rho){dw\over d\rho}+2(\beta+is)w=0.
\ee     
This is familiar. The solution that is well behaved at the origin
is $w(\rho)=\Phi(s-i\beta, 2s+1, -2i\rho)$, where $\Phi(a,b,z)$
is the confluent hypergeometric function [25]. So we have
$$
u_j(\rho)=a_j\rho^s\Phi(s-i\beta, 2s+1, -2i\rho),
\eqno(16{\rm a})$$
where $a_j$ is a constant, and the subscript $j$ of  $u_j$ that is
omitted above has been recovered. Substituting this solution into Eq.
(11a) we have
$$
v_j(\rho)=a_j{s-i\beta\over j+\nu+i\beta'}\rho^s
\Phi(s+1-i\beta, 2s+1, -2i\rho),
\eqno(16{\rm b})$$
where we have used the formula
$$
\left(z{d\over dz}+a\right)\Phi(a,b,z)=a\Phi(a+1,b,z),$$
which can be obtained from other relations given in mathematical
handbooks [25]. It should be remarked that $\psi_j$ may be slightly
singular at the origin when $j=\pm\frac12$. However, the integral
of $\psi_j^\dagger\psi_j$ over any finite volumn converges and the
solution is acceptable. This situation is similar to the case of a
pure Coulomb field [1]. We take
$$
a_j=A2^s(j+\nu+i\beta'){\Gamma(s-i\beta)\over \Gamma(2s+1)}
\exp\left({\beta\pi\over 2}+im\pi-i{s\pi\over 2}+i{\pi\over 4}
\right),
\eqno(17{\rm a})$$
where $m=j-\frac12$ and
$$
A=i\sqrt{2\over\pi k}{1\over\sqrt{k_1+k_2}},
\eqno(17{\rm b})$$
then when $r\to\infty$ we have for the radial wave functions
the asymptotic forms
$$
f_j(r)\to A\sqrt{k_1}\left[i^m\cos\left(kr+\beta\ln 2kr-{m\pi\over2}
-{\pi\over 4}\right)+\frac 12(S_j-1)\exp\left(ikr+i\beta\ln 2kr-
i{\pi\over4}\right)\right], 
\eqno(18{\rm a})$$
$$
g_j(r)\to A\sqrt{k_2}\left[i^m\sin\left(kr+\beta\ln 2kr-{m\pi\over2}
-{\pi\over 4}\right)+\frac 1{2i}(S_j-1)\exp\left(ikr+i\beta\ln 2kr-
i{\pi\over4}\right)\right] 
\eqno(18{\rm b})$$
to the lowest order, where 
\addtocounter{equation}{3}
\bb
S_j=\exp(2i\eta_j)=(j+\nu+i\beta')
{\Gamma(s-i\beta)\over \Gamma(s+1+i\beta)}
\exp(ij\pi-is\pi)
\ee     
and the $\eta_j$'s are phase shifts.
The asymptotic form for $\psi=\sum_j\psi_j$, where the summation is
taken over all $j$, turns out to be
\bb
\psi\to\psi_{\rm in}+\psi_{\rm sc},\quad r\to\infty,
\ee     
where
$$
\psi_{\rm in}={\exp(-im_0\theta)\over\sqrt{k_1+k_2}}
\left(\begin{array}{c}
i\sqrt{k_1}\\
\sqrt{k_2}\end{array}
\right)\varphi_{\rm in},
\eqno(21{\rm a})$$
with
$$
\varphi_{\rm in}=\sum_{m=-\infty}^{+\infty}
i^m\sqrt{2\over \pi kr}\cos\left(kr+\beta\ln 2kr-{m\pi\over2}
-{\pi\over 4}\right)e^{im\theta},
\eqno(21{\rm b})$$
and
\addtocounter{equation}{1}
\bb
\psi_{\rm sc}=\sqrt{\frac ir}\exp(ikr+i\beta\ln 2kr)f(\theta)
{\exp(-im_0\theta)\over\sqrt{k_1+k_2}}
\left(\begin{array}{c}
i\sqrt{k_1}\\
e^{i\theta}\sqrt{k_2}\end{array}
\right),
\ee     
with
\bb
f(\theta)=\sqrt{2\over\pi k}\sum_j\exp(i\eta_j)\sin\eta_je^{im\theta}
=-{i\over\sqrt{2\pi k}}\sum_j(S_j-1)e^{im\theta}.
\ee     
In spite of the presence of the AB potential, the probability
current density associated with a solution $\psi$ is still given by
\bb
{\bf j}=c\psi^\dagger{\bbox\alpha}\psi.
\ee     
As in the case of a pure Coulomb field [1],
one can show that the above $\psi$ represents a correct scattering
solution of which $\psi_{\rm in}$ is an incident wave and
$\psi_{\rm sc}$ the scattered one, $f(\theta)$ is the
scattering amplitude, and the differential cross section is given by
\bb
\sigma(\theta)=|f(\theta)|^2.
\ee            
The choice of $a_j$ in Eq. (17) is thereby proved to be appropriate.
Since $\sum_j e^{im\theta}=2\pi\delta(\theta)$,
we have for $\theta\ne 0$
\bb
f(\theta)
=-{i\over\sqrt{2\pi k}}\sum_j S_je^{im\theta}.
\ee     
This result with $S_j$ given by Eq. (19) is exact. 
However, it is even more difficult to sum up  the
above partial wave series than in the case of a pure Coulomb field.

Let us consider the case with $\gamma^2\ll 1$ and try to work out
a closed result.
For electron-nucleus interaction we have $\gamma\approx Z/137$,
so the above condition means that $Z$ is small, say, $Z<5$.
In this case we may approximately replace $s$ by $|j+\nu|$ in
Eq. (19). Note that $\beta$ also depends on $\gamma$ and we do
not make approximation with it. Thus the result will possess
some nonperturbative features in regard to the parameter
$\gamma$. With the above approximation, $S_j$ is replaced by
$$
S_j^{\rm a}=e^{-i\nu\pi}{\Gamma(m+\nu+1/2-i\beta)\over
\Gamma(m+\nu+1/2+i\beta)}
-i(\beta-\beta')e^{-i\nu\pi}{\Gamma(m+\nu+1/2-i\beta)\over
\Gamma(m+\nu+3/2+i\beta)},\quad (j>0),
\eqno(27{\rm a})$$
$$
S_j^{\rm a}=e^{i\nu\pi}{\Gamma(|m|-\nu+1/2-i\beta)\over
\Gamma(|m|-\nu+1/2+i\beta)}
+i(\beta-\beta')e^{i\nu\pi}{\Gamma(|m|-\nu-1/2-i\beta)\over
\Gamma(|m|-\nu+1/2+i\beta)},
\quad (j<0).
\eqno(27{\rm b})$$
It can be shown that the first term in either equation equals
$\exp(2i\delta_m)$ where $\delta_m$ is the nonrelativistic
phase shift of the $m$th  partial wave, except for $m=0$ when
$-\frac 12<\nu<0$. (The physical reason for the latter disagreement
is not quite clear to us.) The second term is a
relativistic correction which vanishes when $v_{\rm c}/c\to 0$.
Substituting Eq. (27) into Eq. (26), we have an approximate result
for $f(\theta)$:
\addtocounter{equation}{1}
\bb
f^{\rm a}(\theta,\nu)=f_0(\theta,\nu)+f_1(\theta,\nu),
\ee     
where
\begin{eqnarray}
f_0(\theta, \nu)=-{i\over\sqrt{2\pi k}}&&\left[
e^{-i\nu\pi}{\Gamma(1/2+\nu-i\beta)\over \Gamma(1/2+\nu+i\beta)}
F(1,1/2+\nu-i\beta,1/2+\nu+i\beta, e^{i\theta})\right. \nonumber \\
+&&\left.e^{i\nu\pi}{\Gamma(3/2-\nu-i\beta)\over
\Gamma(3/2-\nu+i\beta)}e^{-i\theta}
F(1,3/2-\nu-i\beta,3/2-\nu+i\beta, e^{-i\theta})\right]
\end{eqnarray}     
is the nonrelativistic partial wave result (somewhat different when
$-\frac 12<\nu<0$) and
\begin{eqnarray}
f_1(\theta, \nu)=-{\beta-\beta'\over\sqrt{2\pi k}}&&
\left[e^{-i\nu\pi}{\Gamma(1/2+\nu-i\beta)\over
\Gamma(3/2+\nu+i\beta)}F(1,1/2+\nu-i\beta,3/2+\nu+i\beta,
e^{i\theta})\right.\nonumber \\
-&&\left.e^{i\nu\pi}{\Gamma(1/2-\nu-i\beta)\over
\Gamma(3/2-\nu+i\beta)}e^{-i\theta}
F(1,1/2-\nu-i\beta,3/2-\nu+i\beta, e^{-i\theta})\right]
\end{eqnarray}      
is the relativistic correction. In these equations
$F(a_1, a_2, b_1, z)$ is the hypergeometric function [25].
Using their functional relations it can be shown that
\begin{eqnarray}
f_0(\theta,\nu)=&&-ie^{-i\nu\theta}{\Gamma(1/2-\nu+i\beta)
\Gamma(1/2+\nu-i\beta)\over \Gamma(i\beta)\Gamma(1/2+i\beta)}
{\exp(i\beta\ln\sin^2\theta/2)\over\sqrt{2k}\sin\theta/2}
\nonumber \\
&&-{i\over\sqrt{2\pi k}}\left[e^{i\nu\pi}
{\Gamma(3/2-\nu-i\beta)\over\Gamma(3/2-\nu+i\beta)}-e^{-i\nu\pi}
{\Gamma(-1/2+\nu-i\beta)\over\Gamma(-1/2+\nu+i\beta)}\right]
\nonumber \\
&&\times e^{-i\theta}
F(1,3/2-\nu-i\beta,3/2-\nu+i\beta, e^{-i\theta}),
\end{eqnarray}     
and
\begin{eqnarray}
f_1(\theta,\nu)=&&-\left(1-{\beta'\over\beta}\right)
{\Gamma(1/2+\nu-i\beta)\Gamma(1/2-\nu+i\beta)\over
\Gamma(i\beta)\Gamma(1/2+i\beta)}
{e^{-i\theta/2-i\nu\theta}\exp(i\beta\ln\sin^2\theta/2)\over\sqrt{2k}}
\nonumber \\
&&+{\beta-\beta'\over\sqrt{2\pi k}}\left[e^{i\nu\pi}
{\Gamma(1/2-\nu-i\beta)\over\Gamma(3/2-\nu+i\beta)}+e^{-i\nu\pi}
{\Gamma(-1/2+\nu-i\beta)\over\Gamma(1/2+\nu+i\beta)}\right]
\nonumber \\
&&\times e^{-i\theta}
F(1,1/2-\nu-i\beta,3/2-\nu+i\beta, e^{-i\theta}),
\end{eqnarray}     
where $0\le\theta<2\pi$. Note that these approximate expressions are
well defined for the whole range of $\nu$, and will be employed in the
following discussions for the case when $|\nu|$ is close to
$\frac 12$. Unlike the case with a pure Coulomb
field, here we have additional terms in both $f_0(\theta,\nu)$
and $f_1(\theta,\nu)$, involving the hypergeometric functions.
Thus closed forms are possible only for special $\nu$'s when
the additional terms vanish. This happens when $\nu=0$ and
$\nu=\frac 12$. Since the above discussions are not valid for
$|\nu|$ close to $\frac 12$, we have now a closed result only when
$\nu=0$, or $q\phi/2\pi\hbar c$ takes on integer values. The result is
the same as in the case of a pure Coulomb field: 
\bb
f(\theta)=f^{\rm a}(\theta,0)=
-i{\Gamma(1/2-i\beta)\over \Gamma(i\beta)}
{\exp(i\beta\ln\sin^2\theta/2)\over
\sqrt{2k}\sin\theta/2}\left[1-ie^{-i\theta/2}\sin
{\theta\over 2}\left(1-{\beta'\over\beta}\right)\right].
\ee     
Since the result is more singular than $\delta(\theta)$ when
$\theta\to 0$, the $\delta(\theta)$ term that is dropped above for
$\theta\ne0$ can indeed be dropped everywhere and the above
expression is enough. The differential cross section reads
\bb
\sigma(\theta)
={\beta\tanh\beta\pi\over 2k\sin^2\theta/2}
\left(1-{v_{\rm c}^2\over c^2}\sin^2{\theta\over 2}\right)
={\kappa\tanh(\pi\kappa/\hbar v_{\rm c})\over 2\mu v_{\rm c}^2
\sin^2\theta/2}
\left(1-{v_{\rm c}^2\over c^2}\sin^2{\theta\over 2}\right)
\left(1-{v_{\rm c}^2\over c^2}\right)^{\frac 12},
\ee            
where the first factor in the last expression
is the exact nonrelativistic result,\footnote{In the nonrelativistic
case when $\nu=0$ and $m_0\ne0$,
we got an interference term in the cross section in additional to the
result for a pure Coulomb field, because we excluded the $s$-wave
solution which is slightly singular at the origin [3].
Now it seems that the $s$-wave is
acceptable and that interference term is not necessary, because the
potentials themselves are rather singular at the origin. Indeed,
in the relativistic case, the solutions with $j=\pm 1/2$ are much more
singular, and the singularity is essentially the same as that in a
pure Coulomb field.}
and the subsequent ones are due to the relativistic effect. 
Some remarks similar to those made in Ref. [1]:
First, we have not made any
approximation in regard to the incident velocity, so the result is
valid for high energy collision. It is obvious that the relativistic
correction becomes significant when $v_{\rm c}$
is comparable with $c$.
Second, though the above result holds for small $\gamma$ only, it
involves a nonperturbative factor $\tanh\beta\pi$ (note that
$\beta\propto\gamma$). Thus the result cannot be obtained from
perturbative QED at the tree level.
Moreover, if $m_0$ or $\phi$ is large, perturbative QED seems
not applicable, but the above calculations hold as well.

If $\nu$ is close to $\frac 12$ ($-\frac 12$) but
$|\gamma|<|\frac 12\pm \nu|$, then the above discussions
are still valid. But then the replacement of $S_{-1/2}$ ($S_{1/2}$)
by $S^{\rm a}_{-1/2}$ ($S^{\rm a}_{1/2}$) is a poor approximation.
In this case some corrections are necessary. Since no closed result
is available, we do not discuss it in detail.

Now we turn to the case when $\nu$ is very close to $\frac 12$ 
such that $|\gamma|>|\frac 12-\nu|$ (the typical case is $\nu=
\frac 12$), then the above discussions have to be modified. We will
discuss it in some detail. (The other case with $\nu$ very close to
$-\frac 12$ can be discussed in a similar way and will be
omitted.) The crucial point is that $s$ becomes imaginary 
when $j=-\frac 12$. (A larger $\gamma$ causes the same difficulty
for some more values of $j$.)
So the solutions of Eq. (13) are now given by
\begin{eqnarray}
&&u_{-1/2}^{(1)}(\rho)=a_{-1/2}^{(1)}\exp(i\gamma'\ln\rho)
\Phi(-i\beta+i\gamma',1+2i\gamma',-2i\rho),\nonumber \\
&&u_{-1/2}^{(2)}(\rho)=a_{-1/2}^{(2)}\exp(-i\gamma'\ln\rho)
\Phi(-i\beta-i\gamma',1-2i\gamma',-2i\rho),
\end{eqnarray}  
where
\bb
\gamma'=\sqrt{\gamma^2-(1/2-\nu)^2}.
\ee     
Here the two solutions have essentially the same singularity at the
origin. They oscillate very rapidly and tend to no limit when
$\rho\to 0$. In the conventional opinion of quantum mechanics, such
solutions are not acceptable [26, 27]. One possible way of
resolving the problem is to cutoff the singular potentials (which
represent some idealization) at some small radius. But this causes
some mathematical difficulties. On the other hand, some authors have
attempted to handle such solutions in some convenient way, and for
bound states some proposals have been put forward to choose the
appropriate linear combination of the two solutions [28, 29].
For scattering solutions the situation seems more involved.
Here we will resolve the problem by some simple consideration.
The solutions $v_{-1/2}^{(1)}(\rho)$ and $v_{-1/2}^{(2)}(\rho)$ can
be easily obtained from Eq. (11a) and the above $u$'s. We will not
write them down. We take
\bb
a_{-1/2}^{(1)}=A\left[\beta'+i\left(\frac 12-\nu\right)\right]
{\Gamma(-i\beta+i\gamma')\over\Gamma(1+2i\gamma')}\exp\left(
{\beta\pi\over 2}+{\gamma'\pi\over 2}+i\gamma'\ln 2-i{\pi\over4}
\right),
\ee     
then the asymptotic forms for $f_{-1/2}^{(1)}(r)$ and
$g_{-1/2}^{(1)}(r)$ are given by Eq. (18) with $S_{-1/2}$ replaced
by
\bb
S^{(1)}=\left[\beta'+i\left(\frac 12-\nu\right)\right]
{\Gamma(-i\beta+i\gamma')\over\Gamma(1+i\beta+i\gamma')}
\exp(\gamma'\pi).
\ee     
However, $|S^{(1)}|\ne 1$, so $S^{(1)}$ cannot be expressed as a
phase factor.  We then take $a_{-1/2}^{(2)}$ by replacing $\gamma'$ by
$-\gamma'$ in $a_{-1/2}^{(1)}$,
then the asymptotic forms for $f_{-1/2}^{(2)}(r)$ and
$g_{-1/2}^{(2)}(r)$ are given by Eq. (18) with $S_{-1/2}$ replaced
by $S^{(2)}$ which is obtained by replacing 
$\gamma'$  by $-\gamma'$ in $S^{(1)}$.
As the two solutions above are equally preferable, we take the mean
as the required solution. Then the asymptotic forms for $f_{-1/2}(r)$
and $g_{-1/2}(r)$ are given by Eq. (18) with $S_{-1/2}$ replaced
by
\bb
S=\frac 12[S^{(1)}+S^{(2)}].
\ee     
The solutions with $j\ne -\frac 12$ are all the same as those
given before. So we finally obtain the scattering amplitude (for
$\theta\ne 0$)
\bb
f(\theta)
=-{i\over\sqrt{2\pi k}}[\sum_{j\ne -1/2}S_je^{im\theta}+
Se^{-i\theta}].
\ee     
This is a very complicated result. When $\gamma^2\ll 1$, one can
make approximations as before, but closed results are available
only when $\nu=\frac 12$, or $q\phi/2\pi\hbar c$ takes on half integer
values. We have then
\bb
f(\theta)=f^{\rm a}(\theta,1/2)
-{i\over\sqrt{2\pi k}}(S-S^{\rm a}_{-1/2})e^{-i\theta}.
\ee     
Note that $S^{\rm a}_{-1/2}$ is well defined in Eq. (27b). Since we
have neglected $\gamma^2$ in calculating $f^{\rm a}(\theta,1/2)$,
we can as well neglect it in calculating $S$. It then turns out that
$S=S^{\rm a}_{-1/2}$ to the first order in $\gamma$. So we have
\bb
f(\theta)=f^{\rm a}(\theta,1/2)=
-e^{-i\theta/2}{\beta\Gamma(-i\beta)\over \Gamma(1/2+i\beta)}
{\exp(i\beta\ln\sin^2\theta/2)\over
\sqrt{2k}\sin\theta/2}\left[1-ie^{-i\theta/2}\sin
{\theta\over 2}\left(1-{\beta'\over\beta}\right)\right].
\ee     
The differential cross section reads
\bb
\sigma(\theta)
={\beta\coth\beta\pi\over 2k\sin^2\theta/2}
\left(1-{v_{\rm c}^2\over c^2}\sin^2{\theta\over 2}\right)
={\kappa\coth(\pi\kappa/\hbar v_{\rm c})\over 2\mu v_{\rm c}^2
\sin^2\theta/2}
\left(1-{v_{\rm c}^2\over c^2}\sin^2{\theta\over 2}\right)
\left(1-{v_{\rm c}^2\over c^2}\right)^{\frac 12},
\ee            
where the first factor in the last expression
is the exact nonrelativistic result [3],
and the subsequent ones are due to the relativistic effect. 
The remarks made under Eq. (34) are also applicable here.

Finally we give some results about the bound states. If
$|\gamma|<|\frac 12\pm \nu|$, the solutions can be obtained without
difficulty. The energy levels are
\bb
E_{nj}={\mu c^2 \over [1+\gamma^2/(n+s)^2]^{1/2}},
\ee     
where $n=0,1,2,\ldots$ is a radial quantum number, and $s$ is defined
in Eq. (14), which depends on $j$. The level with $n=0$ is not
degenerate regardless of $\nu$, since one can show that only solutions
with $j>0$ is possible in this case. When
$n>0$, the degeneracy depends on $\nu$. If $\nu\ne 0$, there is no
degeneracy. If $\nu=0$, the energy level depends on $|j|$ rather than
$j$, and solutions with both positive and negative $j$ exist. So the
level is double degenerate. It is remarkable that there are no negative
energy levels. The wave functions are given in terms of confluent
hypergeometric functions. Since the results are complicated we will
not write them down. If $|\gamma|>|\frac 12\pm \nu|$, $s$ will become
imaginary for some $j$. For such values of $j$, special treatment
[28, 29] of the solution is necessary, and the results are rather
involved. We will not go into further details here.

In conclusion, we have calculated the scattering amplitude
and differential cross section for fast
particles with ABC interaction in two dimensions.
Exact results are given in partial wave series in general cases.
Approximate results in closed forms are given in special cases.
Though being approximate, the results exhibit
some nonperturbative features and cannot be obtained from the
lowest-order contribution of perturbative QED. We have also discussed
the bound-state solutions and given some results.

\vskip 1pc

The author is grateful to Professor Guang-jiong Ni for encouragement.
This work was supported by the
National Natural Science Foundation of China.

\end{document}